\documentclass[preprint,showpacs,preprintnumbers,amsmath,amssymb]{revtex4}
\usepackage{array}
\usepackage{graphicx}
\usepackage{dcolumn}
\usepackage{bm}
\usepackage{amsmath, dsfont}
\usepackage{color}

\newcommand{\const}{\mathop{\rm const}\nolimits}

\def\IR{{\mathds{R}}}

\def\vnabla{{\vec{\nabla}}}

\def\vp{{\vec{p}}}
\def\vx{{\vec{r}}}
\def\hx{{\hat{r}}}
\def\vn{{\hat{n}}}

\def\vJ{{\vec{J}}}
\def\vL{{\vec{L}}}
\def\vK{{\vec{K}}}

\def\vA{{\vec{A}}}
\def\vB{{\vec{B}}}

\def\smallover#1/#2{\hbox{$\textstyle\frac{#1}{#2}$}}
\def\IR{{\mathds{R}}}

\def\vA{{\vec{A}}}
\def\vK{{\vec{K}}}
\def\vJ{{\vec{J}}}
\def\vPi{{\vec{\Pi}}}

\def\vv{{\vec{v}}}
\def\vp{{\vec{p}}}
\def\vx{{\vec{x}}}
\def\vn{{\vec{n}}}
\def\va{{\vec{a}}}
\def\vB{{\vec{B}}}
\def\vnabla{{\vec{\nabla}}}
\def\hx{{\frac{\vec{x}}{r}}}
\newtheorem{prop}{Theorem}[section]

\def\beq{\begin{equation}}
\def\eeq{\end{equation}}
\def\beqa{\begin{eqnarray}}
\def\eeqa{\end{eqnarray}}
\def\nn{\nonumber}

\begin{document}

\preprint{arxiv:0908.1204}

\title{Curved manifolds with conserved Runge-Lenz vectors}

\author{\large
 J.-P.~Ngome}
\affiliation{
Laboratoire de Math\'ematiques et de Physique Th\'eorique, 
Universit\'e Fran\c cois-Rabelais de Tours,
F\'ed\'eration Denis Poisson - CNRS
Parc de Grandmont, 37200 Tours, France.
}
\email{ 
ngome-at-lmpt.univ-tours.fr .}

\date{\today}

\begin{abstract}
van Holten's algorithm is used to construct Runge-Lenz-type conserved quantities, induced by Killing tensors, on curved manifolds. For the generalized Taub-NUT metric, the most general external potential  
such that the combined system admits a conserved Runge-Lenz-type vector  is found. In the multicenter case, the subclass of two-center metric exhibits a conserved Runge-Lenz-type scalar.\end{abstract}


\maketitle
\section{Introduction}
\noindent In 1986,  Gibbons and Manton  \cite{GMR} found, in the context of monopole scattering, that the geodesic motion in the Taub-NUT metric \cite{GP-S} admits a Kepler-type dynamical symmetry \cite{FH, GR}. Various generalizations were extensively studied successively \cite{GR2, Japs, LeeLee,Nerses, Valent, Visi, Ballesteros}. More recently, Gibbons and Warnick \cite{GW} considered geodesic motion on hyperbolic space and found a large class of systems admitting such a dynamical symmetry. 

An interesting extension of the Taub-NUT case is provided by the multicenter metrics \cite{GR}. The two-center case exhibits, in particular, an additional conserved quantity, which is quadratic in the momenta \cite{GR, Valent}. 
 
In this paper we present a systematic analysis of metrics with  conserved Runge-Lenz vector.
From now on, we consider the static family of metrics given by
\begin{eqnarray}\displaystyle{
dS^2=f(\vx)\,\delta_{ij}(\vx)\,dx^i\,dx^j+h(\vx)\,\big(dx^4+A_k\,dx^k\big)^2}\,\label{metricTN}\,,
\end{eqnarray}
which contains all previous cases listed above.
In these metrics, $\,f(\vx)\,$ and $\,h(\vx)\,$ are arbitrary real functions and the $1$-form $\,A_k\,$ is the gauge potential of a Dirac monopole of charge  $\,g\,$, $\,A_k\,dx^k=-g\cos\theta\,d\phi\,.$

Our strategy  is that the conservation of the ``vertical'' component of the
momentum allows us to reduce the four-dimensional problem to one in three dimensions, where we have strong candidates for the way these 
symmetries act. Then, the lifting problem can be conveniently solved using the recently proposed technique of Van Holten \cite{vHolten, H-NGI}. 

\section{Geodesic motion}\label{GeoMot}
\noindent The Lagrangian of geodesic motion  on the  $4$-manifold endowed with the metric (\ref{metricTN}) is  
\begin{eqnarray}\displaystyle{
\mathcal{L}=\frac{1}{2}\,f(\vx)\,\delta_{ij}\,\frac{d x^i}{dt}\,\frac{d x^j}{dt}+\frac{1}{2}\,h(\vx)\,\big(\,\frac{d x^4}{dt}+A_k\,\frac{d x^k}{dt}\,\big)^2-U(\vx)}\label{LagrangTN}\,,
\end{eqnarray}
where we also added an external scalar potential, $U(\vx)$, for later convenience.
The canonical momenta conjugate to the coordinates $(x^j,\,x^4)$ read as
\begin{eqnarray}\begin{array}{ll}\displaystyle{
p_j= \frac{\partial \mathcal{L}}{\displaystyle{\partial\big({d x^j}/{dt}\big)}}  =f(\vx)\,\delta_{ij}\,\frac{d x^i}{dt}+h(\vx)\,\big(\,\frac{d x^4}{dt}+A_k\,\frac{d x^k}{dt}\,\big)\,A_j\;,}\ 
\\[14pt]\displaystyle{
p_4= \frac{\partial \mathcal{L}}{\displaystyle{\partial\big({d x^4}/{dt}\big)}} =h(\vx)\,\big(\,\frac{d x^4}{dt}+A_k\,\frac{d x^k}{dt}\,\big)=q}\;.
\end{array}
\end{eqnarray}
 The ``vertical'' momentum, $\;p_4=q\;$, associated with the
 cyclic variable  $x^4$, is conserved and interpreted as conserved electric charge. We, thus, introduce the covariant momentum,
\begin{eqnarray}\displaystyle{
\Pi_j=f(\vx)\,\delta_{ij}\,\frac{d x^i}{dt}=p_j-q\,A_j}\;.\label{DimReduc}
\end{eqnarray}
Geodesic motion on the 4-manifold  projects therefore onto the curved 3-manifold with metric $ \,g_{ij}(\vx)=\,f(\vx)\,\delta_{ij}\,$, augmented with a potential. The Hamiltonian  is
\beqa
\mathcal{H}= \frac{1}{2}\,g^{ij}(\vx)\Pi_{i}\,\Pi_{j}+V(\vx)\quad\hbox{with}\quad V(\vx)=\frac{q^{2}}{2h(\vx)}+U(\vx)\,.
\label{HamVanHolten}
\eeqa
For a particle without spin, the covariant Poisson brackets are given by  \cite{vHolten}
\begin{eqnarray}\displaystyle{
\left\lbrace B,D \right\rbrace = \partial_kB\,\frac{\partial D}{\partial \Pi_k}-\frac{\partial B}{\partial \Pi_k}\,\partial_kD+qF_{kl}\,\frac{\partial B}{\partial \Pi_k}\,\frac{\partial D}{\partial \Pi_l}}\,,
\end{eqnarray}
where $\,\displaystyle{F_{kl}=\partial_kA_l - \partial_lA_k}\,$
is the field strength. Then, the nonvanishing fundamental Poisson brackets are
\beq
\displaystyle{
\left\lbrace x^i,\;\Pi_j\right\rbrace=\delta^i_{j}\ ,
\quad\left\lbrace \Pi_i,\;\Pi_j\right\rbrace =q\,F_{ij} }\,.
\eeq
The Hamilton equations, therefore, read as
\begin{eqnarray}
\dot{x}^i&=&\left\lbrace x^i,\,\mathcal{H}\right\rbrace=g^{ij}(\vx)\,\Pi_j,
\\[5pt]
\dot{\Pi}_{i}&=&\left\lbrace \Pi_i,\,\mathcal{H}\right\rbrace=\displaystyle{
q\,F_{ij} \,\dot{x}^j-\partial_i\displaystyle{V} +\Gamma^k_{ij}\,\Pi_k\,\dot{x}^j}\;. \label{Lorentz}
\end{eqnarray}
Note that the Lorentz equation (\ref{Lorentz}) involves also in addition to the monopole and potential terms, a curvature term which is quadratic in the velocity.

\section{Conserved quantities}\label{algo}
\noindent
Let us recall that
constants of the motion, noted as $\,Q\,$, which are polynomial in the momenta, can be derived following van Holten algorithm \cite{vHolten}. The clue is to expand $\,Q\,$ into a power series of the covariant momentum,
\begin{eqnarray}\displaystyle{
Q= C+C^i\,\Pi_i+\frac{1}{2!}\,C^{ij}\,\Pi_i\Pi_j+\frac{1}{3!}\,C^{ijl}\,\Pi_i\Pi_j\Pi_l+\cdots}\,,
\label{devQbis}
\end{eqnarray}
and to require $Q$ to Poisson-commute with the Hamiltonian augmented with an effective potential, 
$\,\displaystyle{\mathcal{H}=\frac{1}{2}\,\vec{\Pi}^{2}+G(\vx)
}\,$. This yields the series of constraints,
\beqa
\left\lbrace
\begin{array}{lllll} 
C^m\;\partial_m\,G(\vx)=0&(\hbox{order 0}) & \\[9pt]
\partial_n C=q\,F_{nm}\,C^m+C_n^{\;m}\partial_m\,G(\vx) &(\hbox{order 1}) &\\[9pt]
\mathcal{D}_{i}C_{l}+\mathcal{D}_{l}C_i=q\,\left( F_{im}\,C_l^{\;m}+F_{lm}\,C_i^{\;m} \right)+C_{il}^{\;\;k}\partial_k\,G(\vx)&(\hbox{order 2})&\\[9pt]
\mathcal{D}_iC_{lj}+\mathcal{D}_jC_{il}+\mathcal{D}_lC_{ij}=q\,\left( F_{im}\;C_{lj}^{\;\;m}+F_{jm}\;C_{il}^{\;\;m}+ F_{lm}\;C_{ij}^{\;\;m}\right)\\
\qquad\qquad\qquad\qquad\qquad\quad+\quad C_{ijl}^{\;\;\;m}\partial_m\,G(\vx)&(\hbox{order 3})&
\\
\cdots\cdots\,,
\end{array}\right.\label{ConsTraints}
\eeqa
where the zeroth-order constraint can be interpreted as a consistency condition for the effective potential. 
 The expansion can be truncated at a finite order provided 
 some higher-order constraint reduces to a Killing equation,  
\begin{eqnarray}\displaystyle{
\mathcal{D}_{\left(i_1\right.}C_{\left. i_2\;\cdots\;i_n \right)}=0}\;,\label{KillingNAbis}
\end{eqnarray}
where the covariant derivative is constructed with the Levi-Civita connection so that $\,\displaystyle{\mathcal{D}_iC^j =\partial_iC^j +\Gamma^j_{\;ik}\,C^k}\,$. Then, $\,\displaystyle{C_{i_1\cdots i_p}=0}\,$ for all $\,p\,\geqslant\,n\,$ and the constant of motion takes the polynomial form,
\beq
Q=\sum_{k=0}^{p-1}\,\frac{1}{k!}\,C^{i_1\cdots i_k}\,\Pi_{i_1}\cdots\Pi_{i_k}\,.
\eeq

\section{Killing Tensors}\label{KillingSection}
\noindent
van Holten's recipe \cite{vHolten}, presented in section \ref{algo}, is based on Killing tensors~: the conserved angular momentum is associated with a rank-1 Killing tensor (i.e. Killing vector), which generates spatial rotations. Rank-$\,2\,$ Killing tensors lead to conserved quantities quadratic in covariant momentum $\,\vec{\Pi}\,$, etc. In this Section, we discuss  two particular Killing tensors on the $3$-manifold, which carries the metric $\,g_{ij}(\vx)=f(\vx)\,\delta_{ij}\,$. Our strategy is to find conditions for lifting  the Killing tensors, which generate the conserved angular momentum and the Runge-Lenz vector of planetary motion in flat space, respectively, to the ``Kaluza-Klein'' 
$4$-space.

$\bullet\,$ First, we search for a rank-1 Killing tensor which generates ordinary space rotations,
\begin{eqnarray}\displaystyle{C_i=g_{ij}(\vx)\,\epsilon^j_{\;\;kl}\;n^k\,x^l}\,.
\end{eqnarray}
 Requiring $\,\displaystyle{
\mathcal{D}_{\left(i\right.}C_{\left. j \right)}=0}\,$, we obtain the following.
\begin{prop}
On the curved $3$-manifold carrying the metric $\,g_{ij}(\vx)=f(\vx)\,\delta_{ij}$, the rank-1 tensor, $$\,\displaystyle{C_i=g_{ij}(\vx)\,\epsilon^j_{\;\;kl}\;n^k\,x^l}\,,$$ is a Killing tensor when
\begin{eqnarray}
\displaystyle{\left(\vn\times\vnabla\,f(\vx)\right)\cdot\vx=0}\;.
\end{eqnarray}
\label{angularM}
\end{prop}
Note that this condition can be satisfied for some, but not all $\vn$'s.
In the two-center case, for example, it only holds for $\vn$ 
parallel to the axis of the two centers.
When the metric is radial, $f(\vx)=f(r)$,
the gradient is radial, and (\ref{angularM}) holds for all $\vn$.
\vskip2mm

$\bullet\,$ Next, we consider the rank-2 Killing tensor associated with the Runge-Lenz-type conserved quantity
\begin{eqnarray}\displaystyle{C_{ij}=2\,g_{ij}(\vx)\,n_k\,x^k-g_{ik}(\vx)\,n_j\,x^k-g_{jk}(\vx)\,n_i\,x^k}\,,
\label{KillT2}
\end{eqnarray}
inspired by the known flat-space expression. ($\vn$ here is some fixed unit vector).
Then $\,\displaystyle{ \mathcal{D}_{\left( i\right.}C_{\left. j l\right)}}\,$ has to vanish. A tedious calculation yields
\begin{eqnarray}
\begin{array}{lr}\displaystyle{ \mathcal{D}_{\left( i\right.}C_{\left. j l\right)}      =2\,n_{k}\,x^{m}\,\left(\, g_{ij}(\vx)\,\Gamma^{k}_{lm}+g_{il}(\vx)\,\Gamma^{k}_{jm}+g_{jl}(\vx)\,\Gamma^{k}_{im}\,
\right)}-n_{i}\displaystyle{x^{m}\partial_{m} g_{jl}(\vx)}\\[12pt]\quad\quad\quad\quad\quad\quad\quad\quad\quad\quad\quad\quad\quad\displaystyle{-\;n_{j}\,x^{m}\partial_{m}g_{il}(\vx)-n_{l}\,x^{m}\partial_{m}g_{ij}(\vx)\;.
}\end{array}\label{F1}
\end{eqnarray}
Calculating each term on the right hand side of (\ref{F1}), we first obtain
\begin{eqnarray*}\begin{array}{ll}\displaystyle{
n_{i}\,x^{m}\,\partial_{m}g_{jl}(\vx)=f^{-1}(\vx)\,n_{i}\,g_{jl}(\vx)\,x^{m}\,\partial_{m}f(\vx)},
\\[10pt]\displaystyle{
n_{j}\,x^{m}\,\partial_{m}g_{il}(\vx)=f^{-1}(\vx)\,n_{j}\,g_{il}(\vx)\,x^{m}\,\partial_{m}f(\vx)},
\\[10pt]\displaystyle{
n_{l}\,x^{m}\,\partial_{m}g_{ij}(\vx)=f^{-1}(\vx)\,n_{l}\,g_{ij}(\vx)\,x^{m}\,\partial_{m}f(\vx)}\;,\end{array}\label{F2}
\end{eqnarray*}
and next the curvature terms,
\begin{eqnarray*}\begin{array}{cc}\displaystyle{
2\,g_{jl}(\vx)\,n^{k}\,x^{m}\,\Gamma_{im}^{k}=f^{-1}(\vx)\,n_{i}\,g_{jl}(\vx)\,x^{m}\partial_{m}f(\vx)+f^{-1}(\vx)\,n_{m}\,x^{m}\,g_{jl}(\vx)\,\partial_{i}f(\vx)}\\[10pt]\displaystyle{\quad\quad+\quad f^{-1}(\vx)\,n_{k}\,g_{jl}(\vx)\,g^{nk}(\vx)\,g_{im}(\vx)\,x^{m}\,\partial_{n}f(\vx)}\,,
\end{array}\label{F3}
\end{eqnarray*}
\begin{eqnarray*}\begin{array}{cc}\displaystyle{
2\,g_{il}(\vx)\,n^{k}\,x^{m}\,\Gamma_{jm}^{k}=f^{-1}(\vx)\,n_{j}\,g_{il}(\vx)\,x^{m}\partial_{m}f(\vx)+f^{-1}(\vx)\,n_{m}\,x^{m}\,g_{il}(\vx)\,\partial_{j}f(\vx)}\\[10pt]\displaystyle{\quad\quad+\quad f^{-1}(\vx)\,n_{k}\,g_{il}(\vx)\,g^{nk}(\vx)\,g_{jm}(\vx)\,x^{m}\,\partial_{n}f(\vx)}\,,
\end{array}\label{F4}
\end{eqnarray*}
\begin{eqnarray*}\begin{array}{cc}\displaystyle{
2\,g_{ij}(\vx)\,n^{k}\,x^{m}\,\Gamma_{im}^{k}=f^{-1}(\vx)\,n_{l}\,g_{ij}(\vx)\,x^{m}\partial_{m}f(\vx)+f^{-1}(\vx)\,n_{m}\,x^{m}\,g_{ij}(\vx)\,\partial_{l}f(\vx)}\\[10pt]\displaystyle{\quad\quad+\quad f^{-1}(\vx)\,n_{k}\,g_{ij}(\vx)\,g^{nk}(\vx)\,g_{lm}(\vx)\,x^{m}\,\partial_{n}f(\vx)}\,.
\end{array}\label{F5}
\end{eqnarray*}
Inserting into (\ref{F1}), we obtain
\beqa
\mathcal{D}_{\left( i\right.}C_{\left. j l \right)}  
=f^{-1}(\vx)\bigg( g_{\left(ij\right.}\partial_{\left.l\right)}f(\vx)\,n_{m}\,x^{m}- g_{\left(ij\right.}x_{\left.l\right)}\,n^{m}\partial_{m}f(\vx)\bigg)
\,.\nn
\eeqa
Requiring $\,\displaystyle{
\mathcal{D}_{\left( i\right.}C_{\left.j l \right)}=0} \,$ yields the following theorem.
\begin{prop}
On the curved $3$-manifold carrying the metric $\,g_{ij}(\vx)=f(\vx)\,\delta_{ij}$, the tensor $$\,\displaystyle{C_{ij}=2\,g_{ij}(\vx)\,n_k\,x^k-g_{ik}(\vx)\,n_j\,x^k-g_{jk}(\vx)\,n_i\,x^k}$$ is a symmetrical rank-2 Killing tensor associated with the Runge-Lenz-type vector when
\begin{eqnarray}
\displaystyle{
\vn\times\left(\vx\times\vnabla\,f(\vx)\right)=0}\,.
\end{eqnarray}\label{RLenzCond}
\end{prop}\noindent 
It is worth noting that $\mathcal{D}_{k}\,g_{ij}(\vx)=0\,$ due to the compatibility condition of the metric. Therefore, the metric tensor is itself   a symmetrical rank-2 Killing tensor. The associated conserved quantity is the Hamiltonian \cite{GR,vHolten}.

\section{Generalized Taub-NUT metric}\label{GTN}
\noindent
Now we use our Killing tensors to construct conserved quantities 
in the radially symmetric generalized Taub-NUT metrics  (\ref{metricTN}),
\begin{eqnarray}\displaystyle{
dS^2=f(r)\,\delta_{ij}\,dx^i\,dx^j+h(r)\,\big(dx^4+A_k\,dx^k\big)^2}\,\label{radmetricTN}\,.
\end{eqnarray}
Then, the Lagrangian (\ref{LagrangTN})  takes the form,
\begin{eqnarray}\displaystyle{
\mathcal{L}=\frac{1}{2}\,f(r)\,\dot{\vx}^{\,2}+\frac{1}{2}h(r)\,\big(\,\frac{d x^4}{dt}+A_k\,\frac{d x^k}{dt}\,\big)^2-U(r)}\,.\label{LagrangTNG}
\end{eqnarray}
The conserved electric charge and the energy are associated with the cyclic variables $\,x^{4}\,$ and time $\,t\,$, 
\beq
\displaystyle{
q=h(r)\,\big(\,\frac{d x^4}{dt}+A_k\,\frac{d x^k}{dt}\,\big)}\;,\quad\displaystyle{\mathcal{E}=  \frac{\vec{\Pi}^{2}}{2\,f(r)}+\frac{q^2}{2\,h(r)}+U(r)}\,,
\eeq
respectively. The dimensionally reduced Hamiltonian 
can be rearranged as
\begin{eqnarray}
\displaystyle{
\mathcal{H}= \frac{1}{2}\,\vec{\Pi}^{2}+f(r)\,W(r)}
\displaystyle{\quad\hbox{with}\quad
\,W(r)=U(r)+\frac{q^{2} }{2\,h(r)}+ \frac{\mathcal{E} }{f(r)}-\mathcal{E}}\,.\label{FormPot1}
\end{eqnarray}

$\bullet$ First, we look for conserved angular momentum, linear in the covariant momentum.  $\,C_{ij}=C_{ijk}=\cdots =0\,$ so that (\ref{ConsTraints}) reduces to
\beqa\left\lbrace \begin{array}{lll} 
 C^m\;\partial_m\,\big(f(r)\,W(r)\big)=0&(\hbox{order 0})& 
\\[8pt]
\partial_n C=q\,F_{nm}\,C^m &(\hbox{order 1})& 
\\[8pt]
\mathcal{D}_{i}C_{l}+\mathcal{D}_{l}C_i=0&(\hbox{order 2})\,.&
\label{SystOrder1}
\end{array}\right.
\eeqa
The $3$-metric now satisfies Theorem (\ref{angularM}). The second- and the first-order constraints yield
\beq
\displaystyle{
C_i=g_{im}(r)\,\epsilon^m_{\;\;\;nk}\;n^n\;x^k }
\quad\hbox{and}\quad\displaystyle{
C=-qg\,n_k\,\frac{x^k}{r}}\,,
\eeq
respectively.
The zeroth-order consistency condition in (\ref{SystOrder1})  is satisfied for an arbitrary radial effective potential, providing us with the conserved angular momentum which involves the  typical monopole term,
\begin{eqnarray}
\displaystyle{\vJ=\vx\times\vPi-qg\,\hx}\,.
\label{ANGTN}
\end{eqnarray}
Let us now turn to quadratic conserved quantities. We have $\,C_{ijk}=C_{ijkl}=\nobreak\cdots =0\,$ which implies the constraints,
\begin{eqnarray}\left\lbrace \begin{array}{llll} 
C^m\;\partial_m\,\big(f(r)\,W(r)\big)=0&(\hbox{order 0})& 
\\[10pt]
\partial_n C=q\,F_{nm}\,C^m+C_{n}^{\;m}\,\partial_m\,\big(f(r)\,W(r)\big)&(\hbox{order 1}) & 
\\[10pt]
\mathcal{D}_{i}C_{l}+\mathcal{D}_{l}C_i=q\,\left( F_{im}\,C_l^{\;m}+F_{lm}\,C_i^{\;m} \right)&(\hbox{order 2})&
\\[10pt]
\mathcal{D}_iC_{lj}+\mathcal{D}_jC_{il}+\mathcal{D}_lC_{ij}=0&(\hbox{order 3})\,.&
\end{array}\right.\label{SystOrder2}
\end{eqnarray}

$\bullet$ We first take, $\,C_{ij}=g_{ij}(r)\,$, as a rank-2 Killing tensor, to deduce from the second-order equation of (\ref{SystOrder2}) that $\,C_i =0\,$. The zeroth-order and the first-order  consistency relations are satisfied by any radial effective potential $\,C=f(r)\,W(r)\,$.  The conserved quantity 
associated to this Killing tensor is therefore the Hamiltonian,
\begin{eqnarray}
\mathcal{H}= \frac{1}{2}\vec{\Pi}^{2}+f(r)\,W(r)\,.
\end{eqnarray}

$\bullet$ Next, we search for a Runge-Lenz-type vector. For this, we have to solve the system of Eq.(\ref{SystOrder2}) with the rank-2 Killing tensor 
\begin{eqnarray}\displaystyle{ C_{ij}= 2\,g_{ij}(r)\,n_k\,x^k-g_{ik}(r)\,n_j\,x^k-g_{jk}(r)\,n_i\,x^k}
\label{RL1}
\end{eqnarray}
inspired by its form in the Kepler problem.
The radial metric satisfies Theorem (\ref{RLenzCond}).  Solving the second-order constraint of (\ref{SystOrder2}), we get
\begin{eqnarray}\displaystyle{ C_{i}= \frac{q\,g}{r}\,g_{im}(r)\,\epsilon^{m}_{\;\;\;jk}\,n^j\,x^k}\,.\label{RL2}
\end{eqnarray} 
Next, inserting (\ref{RL1}) and (\ref{RL2}) into the first-order constraint of (\ref{SystOrder2}) leaves us with
\begin{eqnarray*}
\displaystyle{
\partial_j C=\left(\frac{\big(f(r)\,W(r)\big)'}{r}+\frac{q^2g^2}{r^4}\right)x_j\,n_k\,x^k-\left(r\big(f(r)\,W(r)\big)'+\frac{q^2g^2}{r^2} \right)n_j}\,.\end{eqnarray*}
The integrability condition of this equation is obtained 
by requiring the vanishing of the commutator,
\begin{eqnarray}\displaystyle{
\left[\,\partial_i\,,\,\partial_j\, \right]C=0\quad\Longrightarrow\quad
\Delta\left(f(r)\,W(r)- \frac{q^2g^2}{2r^2}\right)=0 }\,.
\label{Laplacecond}
\end{eqnarray}
Thus, the bracketed quantity must satisfy the 
\emph{Laplace equation}. So the zeroth-order equation is identically satisfied. Consequently, a Runge-Lenz-type conserved vector does exist when the radial effective potential is equal to
\begin{eqnarray}\displaystyle{
f(r)\,W(r)=\frac{q^2g^2}{2r^2}+\frac{\beta}{r}+\gamma}\quad\hbox{with}\quad\displaystyle{\beta, \gamma\;\in\;\IR }\,.\label{FormPot2}
\end{eqnarray}
The results (\ref{FormPot1}) and (\ref{FormPot2}) allow us to enunciate a theorem equivalent to that of Gibbons and Warnick \cite{GW}.
\begin{prop}
For the generalized Taub-NUT metric (\ref{radmetricTN}), the most general potentials $\,U(r)\,$ admitting a Runge-Lenz-type conserved vector are given by
\begin{eqnarray}
U(r)=\left(\frac{q^2g^2}{2r^2}+\frac{\beta}{r}+\gamma
 \right)\frac{1}{f(r)} -\frac{q^2}{2\,h(r)}+\mathcal{E}\,,
 \label{pot37}
\end{eqnarray}
\label{PotCond}
where $q$ and $g$ are the particle and the monopole charge, respectively, 
$\beta$ and $\gamma$ are free constants, and $\mathcal{E}$
is the fixed energy [cf. (\ref{FormPot1})].
\end{prop}
Inserting (\ref{FormPot2}) into the first-order constraint of (\ref{SystOrder2}) yields, 
$
\displaystyle{
\partial_n C=\frac{\beta}{r}\,n_n-\frac{\beta}{r^3}\,n_k\,x^k\,x_n }\,,
$
 which is solved by
\beq
\displaystyle{C=\frac{\beta}{r}\,n_k\,x^k }\,.\label{RL3}
\eeq
Collecting the results (\ref{RL1}), (\ref{RL2}) and (\ref{RL3}) yield the conserved Runge-Lenz-type vector,
\begin{eqnarray}\displaystyle{
\vec{K}=\vec{\Pi}\times\vec{J}+\beta\,\frac{\vec{x}}{r}}\,.\label{RLTN}
\end{eqnarray}
Due to the simultaneous existence of the conserved angular momentum (\ref{ANGTN}) and the conserved Runge-Lenz vector (\ref{RLTN}), the motions of the particle are confined to conic sections \cite{FH}. Our class of metrics which satisfy Theorem \ref{PotCond} includes the following.

\begin{enumerate}
\item
The original Taub-NUT case \cite{GP-S}  with no external $U(r)=0$,
\beq
f(r)=\frac{1}{h(r)}
=1+\frac{4m}{r},
\label{TNcase}
\eeq
where $m$ is real \cite{GR,FH}. (Monopole scattering corresponds to $m=-1/2$ \cite{GMR,FH}). We obtain, for $\,\gamma=q^2/2-\mathcal{E}\,$  and charge $\,g=\pm 4m\,$, the conserved Runge-Lenz vector,
\beq
\vec{K}=\vPi\times\vec{J}-4m\left(\mathcal{E}-q^{2}\right)\hx\,.\label{KKRL}
\eeq
\item  Lee and Lee \cite{LeeLee} argued that for monopole
scattering with independent
components of the Higgs expectation values, the geodesic
Lagrangian (\ref{LagrangTN}) should be replaced by $L\to L-U(r)$,
where
\beq
U(r)=\frac{1}{2}\,\frac{a^{\;\,2}_{0}}{1+\displaystyle\frac{4m}{r}}\ .
\label{LLpot}
\eeq
It is now easy to see that this addition merely shifts the
value in the brackets in (\ref{Laplacecond}) by a constant (a shift of $a^{\;\,2}_{0}/2$ in the energy ) so that Laplace's equation is still satisfied. So the previously
found Runge-Lenz vector (\ref{KKRL}) is still valid.
\item
The metric associated with winding strings \cite{GR2}
\beq
f(r)=1,
\qquad 
h(r)=\frac{1}{\big(1-\displaystyle\frac{1}{r}\big)^2}\,.
\eeq
For charge $\,g=\pm 1\,$, we deduce from Theorem \ref{PotCond},
\beq\displaystyle{\left(\beta+q^{2}\right)-r\,\left(U(r)-\gamma+\frac{q^{2}}{2}-\mathcal{E}\right)=0}\,,\nn
\eeq
so that for the fixed energy, $\,\mathcal{E}=q^{2}/2-\gamma+U(r)\,$, the conserved Runge-Lenz vector reads as
\beq\displaystyle{
\vec{K}=\dot{\vx}\times\vec{J}-q^{2}\,\hx}\,.
\eeq
\item The extended Taub-NUT metric \cite{Japs}
\begin{eqnarray}\begin{array}{cc}\displaystyle{
f(r)= b+\frac{a}{r},
\quad 
h(r)= \frac{a\,r+b\,r^2}{1+d\,r+c\,r^2}}\,,
\end{array}
\label{japexp}
\end{eqnarray}
with $a,\, b,\, c,\, d,\,\in\IR $ \footnote{
The oscillator-type system of Ref. \cite{Japs} is obtained for a suitable modification.}\,.
With the choices $\,U(r)=0\,$ and charge $\,g=\pm 1\,$,  Theorem \ref{PotCond} requires
\begin{eqnarray*}\displaystyle{
-r\,f(r)\,\mathcal{E}+\frac{r\,f(r)}{h(r)}\,\frac{q^{2}}{2}-\frac{q^{2}}{2\,r}-\gamma\,r}=\beta=\const\,.
\end{eqnarray*}
Inserting here (\ref{japexp}) yields
$$
\Big(-a\,{\cal E}+\frac{1}{2}d\,q^2-\beta\Big)+
r\Big(-b\,{\cal E}+\frac{1}{2}c\,q^2-\gamma \Big)=0,
$$
which holds when $
\beta=-a\,{\cal E}+\frac{1}{2}d\,q^2\,$ and $\,\gamma=-b\,{\cal E}+\frac{1}{2}c\,q^2.
$
Then we get the conserved Runge-Lenz vector,
\beq
\vec{K}=\vPi\times\vec{J}-\left(a\,
\mathcal{E}-\frac{1}{2}\,d\,q^{2}\right)\hx\,.
\eeq

\end{enumerate}

\section{The Multi-center metric}\label{2center}

\noindent
Let us consider a particle moving in the Gibbons-Hawking space \cite{GH} which generalizes the Taub-NUT space. The Lagrangian function associated with this dynamical system is given by
\begin{eqnarray*}\displaystyle{
\mathcal{L}=\frac{1}{2}\,f(\vx)\,\dot{\vx}^{\,2}+\frac{1}{2}f^{-1}(\vx)\,\big(\,\frac{dx^4}{dt}+A_k\,\frac{dx^k}{dt}\,\big)^2-U(\vx)}\end{eqnarray*}
so that the functions $\,f(\vx)\,$ obey the ``self-dual'' ANSATZ 
$\,\displaystyle{
\vnabla\,f(\vx)=\pm\vnabla\times\vA}\,$ \cite{GH}. Hence, we obtain the three-dimensional Laplace equation,
\begin{eqnarray*}\displaystyle{
\Delta\,f(\vx)=0}\,,
\end{eqnarray*}
the most general solution of which is given by
\begin{eqnarray}\displaystyle{
f(\vx)=f_{0}+\sum_{i=1}^{N}\frac{m_{i}}{\vert\vx-\va_{i}\vert}}\quad\hbox{with}\quad(f_{0}\,,\;m_{i})\in \IR ^{N+1}\,.
\end{eqnarray}
The multicenter metric admits multi-NUT singularities so that the position of the \textit{ith} NUT singularity with the charge $\,m_{i}\,$ is $\,\va_{i}\,$. We can remove these singularities provided we choose all the NUT charges equal,  $\,\displaystyle{m_{i}=\frac{g}{2}}\,$. In this case, the cyclic variable $\,x^{4}\,$ is periodic with the range $\,\displaystyle{0 \leq x^{4}\leq 4\pi\,\frac{g}{N}}\,$.
We are interested in the projection of the motion of the particle onto the curved 3-manifold described by the metric,
\begin{eqnarray}\displaystyle{
g_{jk}(\vx)=\left(
f_{0}+\sum_{i=1}^{N}\frac{m_{i}}{\vert\vx-\va_{i}\vert}\right)\,\delta_{jk}}\,.\label{MCmetric}
\end{eqnarray}
The Hamiltonian is given by
\beq
\displaystyle{
\mathcal{H}= \frac{1}{2}\vec{\Pi}^{2}+f(\vx)\,W(\vx)}\;,\quad\displaystyle{
\,W(\vx)=U(\vx)+\frac{q^{2}}{2}f(\vx)+ \mathcal{E}f^{-1}(\vx)-\mathcal{E}}\,.
\eeq
For simplicity, we limit ourselves to a discussion of the two-center metrics, 
\beq\displaystyle{
f(\vx)=f_{0}+\frac{m_{1}}{\vert\vx-\va\vert}+\frac{m_{2}}{\vert\vx+\va\vert}}\,,
\eeq
which include, as special regular cases, those listed in 
Table \ref{Table1}.
\begin{table}\caption{Examples of two-center metrics.}
\begin{tabular}{| l | c | r |}
  \hline\hline
  $\,f_{0}\,$ & $\,N\,$ & Type of Metric \\
  \hline\hline
  0 & 1 & ($m_{1}$ or $m_{2}$ = 0) Flat space\;\\ \hline
  1 & 1 & ($m_{1}$ or $m_{2}$ = 0) Taub-NUT\, \\ \hline
  0 & 2 & Eguchi-Hanson\, \\\hline
  1 & 2 & Double Taub-NUT\,.\\
  \hline\hline 
  \end{tabular}
    
  \label{Table1}
\end{table}

$\bullet$ Finding a conserved quantity linear in the covariant momentum requires solving the system,
\beqa\left\lbrace\begin{array}{lll}
C^m\;\partial_m\,\big(f(\vx)\,W(\vx)\big)=0 &(\hbox{order 0})& 
\\[8pt]
\partial_n C=q\,F_{nm}\,C^m &(\hbox{order 1})& 
\\[8pt]
\mathcal{D}_{i}C_{l}+\mathcal{D}_{l}C_i=0 &(\hbox{order 2})\,.& 
\label{SystMCOrder1}
\end{array}\right.
\eeqa
From Theorem \ref{angularM}, we deduce  the rank-1 Killing tensor satisfying the second-order constraint of (\ref{SystMCOrder1}),
\begin{eqnarray}\displaystyle{
C_i=g_{im}(\vx)\,\epsilon^m_{\;\;\;l k}\;\frac{a^l}{a}\,x^k }\,,\label{Killing1MC}
\end{eqnarray}
which generates rotational symmetry around the axis of the two centers. Next,  injecting both (\ref{Killing1MC}) and the magnetic field of the two centers,
\beq
\displaystyle{\vB=m_{1}\,\frac{\vx-\va}{\vert\vx-\va\vert^{3}}+m_{2}\,\frac{\vx+\va}{\vert\vx+\va\vert^{3}}}\label{2CenterMagneticField}
\eeq
 into the first-order equation of (\ref{SystMCOrder1}) yield  
\beq\displaystyle{C=-q\,\left( m_1\,\frac{\vx-\va}{\vert \vx-\va\vert}+m_2\,\frac{\vx+\va}{\vert \vx+\va\vert}\right)\cdot\frac{\va}{a} }\;.
\eeq
Finally we obtain, as conserved linear quantity, the projection of the angular momentum on the axis of the two centers,
\begin{eqnarray}\displaystyle{\mathcal{J}_a=\mathcal{L}_{a}-q\left( m_1\,\frac{\vx-\va}{\vert \vx-\va\vert}+m_2\,\frac{\vx+\va}{\vert \vx+\va\vert}\right)\cdot\frac{\va}{a}}\quad\hbox{with}\quad\displaystyle{\mathcal{L}_{a}=\left(\vx\times\vPi\right)\cdot\frac{\va}{a}}\,,
\end{eqnarray}
which is consistent with the axial symmetry of the two-center metric.

\vskip2mm
Now we study quadratic conserved quantities,
$\,\displaystyle{Q=C+C^i\,\Pi_i+\frac{1}{2}\,C^{ij}\,\Pi_i\Pi_j}\,$. Putting $\,C_{ijk}=C_{ijkl}=\nobreak\cdots =0\,$, leaves us with,
\beqa\left\lbrace\begin{array}{llll}
C^m\;\partial_m\,\big(f(\vx)\,W(\vx)\big)=0&(\hbox{order 0})&\\[8pt]
\partial_n C=q\,F_{nm}\,C^m+C_{n}^{\;m}\,\partial_m\,\big(f(\vx)\,W(\vx)\big)&(\hbox{order 1})&
\\[8pt]
\mathcal{D}_{i}C_{l}+\mathcal{D}_{l}C_i=q\,\left( F_{im}\,C_l^{\;m}+F_{lm}\,C_i^{\;m} \right)&(\hbox{order 2})&
\\[8pt]
\mathcal{D}_iC_{lj}+\mathcal{D}_jC_{il}+\mathcal{D}_lC_{ij}=0&(\hbox{order 3})\,.&
\end{array}\right.\label{SystMCOrder2}
\end{eqnarray}

$\bullet$ Let us first consider the reducible rank-2 Killing tensor,
\begin{eqnarray}\displaystyle{
C_{ij}=\frac{2}{a^2}\,g_{im}(\vx)\,g_{jn}(\vx)\,\epsilon^m_{\;\;\;l k}\,\epsilon^n_{\;\;\;p q}\,a^l\,a^p\,x^k\,x^q+\frac{2}{a^2}\,g_{il}(\vx)\,g_{jm}(\vx)\,a^l\,a^m}\,,\label{KillTensor}
\end{eqnarray}
which is a symmetrized product of Killing-Yano tensors. $\,\displaystyle{C_{i}=g_{im}(\vx)\,\epsilon^m_{\;\;\;l k}\;\frac{a^l}{a}\,x^k}\,$ generates rotations around the axis of the two centers and $\,\displaystyle{\widetilde{C}_j=g_{jm}(\vx)\,\frac{a^m}{a} }\,$ generates spatial translation along the axis of the two centers. We inject (\ref{KillTensor}) in the second-order constraint of (\ref{SystMCOrder2}). This yields
\begin{eqnarray}\displaystyle{ C_{i}= -\frac{2\,q}{a^{2}}\,g_{im}\,\epsilon^m_{\;\;\;jk}\,a^j\,x^k\,a_{l}\left(m_1\,\frac{x^{l}-a^{l}}{\vert \vx-\va\vert}+m_2\,\frac{x^{l}+a^{l}}{\vert \vx+\va\vert}\right)}\,.
\end{eqnarray} 
For null effective potential, we solve the first-order constraint with
\beq
\displaystyle{
C=\frac{q^2}{a^{2}}\,\left(m_1\,\frac{\left(x^{l}-a^{l}\right)}{\vert \vx-\va\vert}\,a_{l}+m_2\,\frac{\left(x^{l}+a^{l}\right)}{\vert \vx+\va\vert}\,a_{l}\right)^{2}}\,,
\eeq
so that we obtain the \emph{square of the projection of the angular momentum onto the axis of the two centers, plus 
a squared component along the axis of the two centers of the covariant momentum}, 
\begin{eqnarray}\displaystyle{
Q=\mathcal{J}_a^{2}+\Pi^2_a}\,.
\end{eqnarray}
As expected, this conserved quantity is not really a new constant of the motion \cite{GR,Valent}.

$\bullet$ Now we turn to the Runge-Lenz-type conserved quantity. Applying Theorem \ref{RLenzCond} to the two-center metric, we obtain
\beq\displaystyle{
\vn\times\left(\vx\times\vnabla\,f\right)=\left(\frac{m_2}{\vert\vx+\va\vert^3}-\frac{m_1}{\vert\vx-\va\vert^3}\right)\left(\vx\times\va \right)\times\vn}\,.
\eeq
The previous term
vanishes only when  $\,\vx\,$ is  parallel to $\,\va\,$, or
when
\begin{eqnarray}\displaystyle{
\frac{m_2}{\vert\vx+\va\vert^3}-\frac{m_1}{\vert\vx-\va\vert^3}=0}\,.\label{SphereMotion}
\end{eqnarray}
Assuming that both charges are positive $\,m_{1}>0\,$, $\,m_{2} > 0\,$ and $\,\va=\left(a_{1},\,a_{2},\,a_{3}\right)\,$, thus the Eq.(\ref{SphereMotion}) becomes 
\begin{eqnarray}\begin{array}{cc}\displaystyle{
\left(x-a_{1}\,\rho\right)^{2}+\left(y-a_{2}\,\rho\right)^{2}+\left(z-a_{3}\,\rho\right)^{2}=    a^{2}\,\left(\rho^{2}-1\right)}\\[10pt]\displaystyle{
\hbox{with}\quad\rho = \frac{m_1^{2/3}+m_2^{2/3}}{m_2^{2/3}-m_1^{2/3}}}\;.
\end{array}
\end{eqnarray}
We recognize here the equation of a 2-sphere of center $\,\va\,\rho\,$ and radius $\,\displaystyle{R = a\,\sqrt{\rho^{2}-1}}\,$, noted as $\,\mathcal{S}^{2}\,$.

Let us first check that the motions 
can be consistently confined onto the $2$-sphere $\,\mathcal{S}^{2}\,$.  We assume that the 
initial velocity is tangent to $\,\mathcal{S}^{2}\,$ and, using the equations of motion, we verify that at  time  $t+\delta t\,$  the velocity remains tangent to $\,\mathcal{S}^{2}\,$. Thus we write
\beq
\vv(t_{0}+\delta t)= \vv_{0}+ \delta t\;\dot{\vv}_{0}\quad\hbox{with $ \vv_{0}=\vv(t_{0})$ tangent to $\,\mathcal{S}^{2}\,$}.\label{Acc}
\eeq
The equations of motion  in the effective scalar potential (\ref{FormPot3}) read as
\beq
\dot{\vPi}=q\,\vv\times\vB-\vnabla\left(f(\vx)W(\vx)\right)-\frac{v^{2}}{2}\left(\frac{m_1}{\vert\vx-\va\vert^3}+\frac{m_2}{\vert\vx+\va\vert^3}\right)\vx\,.
\eeq
Injecting the expressions of the magnetic field of the two-center (\ref{2CenterMagneticField}) and the effective potential (\ref{FormPot3}), we obtain
\beq
\dot{\vPi}_{0}=\left(\frac{m_1}{\vert\vx-\va\vert^3}+\frac{m_2}{\vert\vx+\va\vert^3}\right)\left[q^{2}\left(\frac{m_1}{\vert\vx-\va\vert}+\frac{m_2}{\vert\vx+\va\vert}\right)-\frac{v^{\;2}_{0}}{2}+\beta\right]\vx_{0}=f(\vx)\,\dot{\vv}_{0}\,.
\eeq
Thus  
$\vv(t_{0}+\delta t)$
in (\ref{Acc}) becomes
$$
 \vv_{0}+ \gamma\,\delta t\;\vx_{0}\,,
\quad
\gamma= f^{-1}(\vx)\left(\frac{m_1}{\vert\vx-\va\vert^3}+\frac{m_2}{\vert\vx+\va\vert^3}\right)\left(q^{2}\left(\frac{m_1}{\vert\vx-\va\vert}+\frac{m_2}{\vert\vx+\va\vert}\right)-\frac{v^{\;2}_{0}}{2}+\beta\right),
$$
 where $\vv_{0}$ and $\vx_{0}$ are tangent to the $2$-sphere $\,\mathcal{S}^{2}\,$.
Thus the velocity remains tangent to $\,\mathcal{S}^{2}\,$.

Having shown the consistency, we obtain the following theorem.
\begin{prop}
In the curved 3-manifold carrying the 2-center metric (\ref{MCmetric}), a Runge-Lenz-type conserved quantity does exist only for a particle moving along the axis of the two centers or for motions confined on the 2-sphere of radius $\,\displaystyle{R = a\,\sqrt{\rho^{2}-1}}\,$ centered at $\,\va\,\rho\,$ $(\,m_{1},\,m_{2}>0)\,$. In the Eguchi-Hanson case $\,(m_{1}=m_{2})\,$, the 2-sphere is replaced by the median plane of the two centers.\label{SphereTh}
\end{prop}\noindent
From now on, 
we consider motions confined to a 2-sphere. 
Replacing the rank-2 Killing tensor by its expression (\ref{KillT2}) in the second-order constraint of (\ref{SystMCOrder2}), we obtain
\begin{eqnarray}\displaystyle{ C_{i}= \frac{q}{a}\,g_{im}\,\epsilon^m_{\;\;\;jk}\,a^j\,x^k\,\left(\frac{m_1}{\vert \vx-\va\vert}+\frac{m_2}{\vert \vx+\va\vert}\right)}\,,\label{RL2MC}
\end{eqnarray} 
where the only component of $\,\vn\,$ is along  the axis $\,\displaystyle{\va}/{a}\,$. The final step is to solve the first-order constraint of (\ref{SystMCOrder2}). Following (\ref{FormPot2}), the clue is to choose
\begin{eqnarray}
f(\vx)\,W(\vx)=
\frac{q^{2}}{2}\left( \frac{m_1}{\vert \vx-\va\vert}+\frac{m_2}{\vert \vx+\va\vert} \right)^{2}
+\beta\left( \frac{m_1}{\vert \vx-\va\vert}+\frac{m_2}{\vert \vx+\va\vert} \right)+\gamma
\label{FormPot3}
\end{eqnarray}
with $\beta, \gamma\;\in\;\IR$.
The leading coefficient of the effective potential cancels the obstruction due to the magnetic field of the two centers; and the remaining part in the right hand side leads to
\begin{eqnarray}\displaystyle{C=\beta\,\left( m_1\,\frac{\vx-\va}{\vert \vx-\va\vert}+m_2\,\frac{\vx+\va}{\vert \vx+\va\vert}\right)\cdot\frac{\va}{a} 
}\,.\label{RL3MC}
\end{eqnarray}
Collecting our results yields
\beq\displaystyle{
K_{a}=\left(\vPi\times\vJ\right)\cdot\frac{\va}{a}+\frac{\beta}{q}\,\left(\mathcal{L}_{a}-\mathcal{J}_{a}\right)
}\,,\label{RLScalar}
\eeq
which is indeed a conserved Runge-Lenz-type scalar for particle moving on the 2-sphere of center positioned at $\,\va\,\rho\,$ and radius $\,\displaystyle{R = a\,\sqrt{\rho^{2}-1}}\,$, combined with the effective potential (\ref{FormPot3}). This potential indeed satisfies the consistency condition given by the zeroth-order constraint of (\ref{SystMCOrder2}).
£
\section{Killing-St\"ackel Tensors on extended manifold}\label{LiftedKillingTensor}
\noindent
Let us consider the geodesic motion of a particle before  dimensional reduction (\ref{DimReduc}). The particle evolves on the extended 4-manifold carrying the metric $\,g_{\mu\nu}(x)\,$. A rank-2 Killing-St\"ackel tensor on this curved 4-manifold is a symmetric tensor, $\,C_{\mu\nu}\,$, which satisfies
\beq\mathcal{D}_{\left( \lambda\right.}C_{\left.\mu\nu\right)} =0\,.
\eeq
For the Killing-St\"ackel tensor generating the Runge-Lenz-type conserved quantity, the degree-2 polynomial function in the canonical momenta $p_{\mu}\,$ which are associated with the local coordinates $\,x^{\mu}\,$,
\beq
K=\frac{1}{2}\,C^{\mu\nu} \, p_{\mu} \, p_{\nu}\quad(\mu,\,\nu=1,\cdots,4)
\eeq
is preserved along geodesics. Then, the lifted Killing-St\"ackel tensor on the 4-manifold, which directly yields the Runge-Lenz-type conserved quantity, is written as
\beq
\displaystyle{
C^{\mu\nu} =  \left(\begin{array}{cc} C^{i\,j}  & \quad C^{i\,4}\\[10pt]C^{4\,j}  & \quad C^{4\,4} \end{array}\right)\quad (i,\,j=1,2,3)}\,.\label{LKT1}
\eeq
The tensor $\,C_{i\,j} \,$ is, therefore, the rank-2 Killing tensor on the dimensionally reduced curved 3-manifold carrying the metric $ \,g_{ij}(\vx)=\,f(\vx)\,\delta_{ij}\,$, which generates the Runge-Lenz-type conserved quantity along the projection of the geodesic motion onto the curved 3-manifold. The off-diagonal and the diagonal contravariant components are, respectively,
\beq
C^{i\,4}=C^{4\,i}=\frac{1}{q}\,C^{i}-C^{i}_{\;\;k}\,A^{k}\quad\hbox{and}\quad
C^{4\,4}=\frac{2}{q^{2}}\,C-\frac{2}{q}C_{k}\,A^{k}+C_{jk}\,A^{j}A^{k}\,.
\eeq
The term $\,A^{k}\,$ represents the component of the vector potential of the magnetic field. In the case of the generalized Taub-NUT metric, the terms $\,C\,$ and $\,C_{k}\,$ are the results (\ref{RL3}) and (\ref{RL2}) of the first- and the second-order constraints of (\ref{SystOrder2}), respectively. In the case of the two-center metric, $\,C\,$ and $\,C_{k}\,$ are given by the results (\ref{RL3MC}) and (\ref{RL2MC}), respectively.

$\bullet$ Let us now consider a particle in the gravitational potential, $\,\displaystyle{V(r)=-\frac{m_{0}\,G_{0}}{r}}\,$, described by the Lorentz metric \cite{DGH},
\beq
dS^{2}=d\vx^{\,2}+2\,dx^{4}dx^{5}-2\,V(r)\left(dx^{5}\right)^{2}\,.\label{BargMetric}
\eeq
The variable $\,x^{5}=t\,$ is the non-relativistic time  and $\,x^{4}\,$ the vertical coordinate. Rotations, time translations and ``vertical'' translations generate as conserved quantities the angular momentum $\,\vL\,$, the energy and the mass $\,m\,$, respectively. The Runge-Lenz-type conserved quantity, along null geodesics of the 5-manifold described by the metric (\ref{BargMetric}),
\beq
K=\frac{1}{2}\,C^{ab} \, p_{a} \, p_{b}\quad\hbox{with}\quad a,\,b,\,c=1,\cdots, 5\
\eeq
 is derived from the trace-free rank-2 Killing-St\"ackel tensor \cite{DGH},
\beq
C^{ab}=\left(\frac{\hat{\eta}}{g^{c}_{\;\;c}}\right)g^{ab}-\eta^{ab}\quad\hbox{with}\quad\hat{\eta}=\eta^{ab}g_{ab}\,.\label{LlightKillingTensor}
\eeq
For some $\,\vn\,\in\IR^{3}\,$, the nonvanishing contravariant components of $\,\eta\,$ are given by
\beq
\eta^{ij}=n^{i}\,x^{j}+n^{j}\,x^{i}-\hat{\eta}\,\delta^{ij}\quad\hbox{and}\quad\eta^{45}=\eta^{54}=\hat{\eta}=n_{i}\,x^{i}\,.
\eeq
A calculation of each matrix element of the Killing tensor (\ref{LlightKillingTensor}) leads to $C^{ab} $ whose only nonvanishing components
are,
\beq
\displaystyle{
C^{ij} =  2\,\hat{\eta}\,\delta^{ij}-n^{i}\,x^{j}-n^{j}\,x^{i}\quad\hbox{and}\quad C^{44} =2\,\hat{\eta}\,V(r)\,.}\label{LlightKillingTensor2}
\eeq
The associated Runge-Lenz-type conserved quantity reads as
\beq
\vK\cdot\vn=\left(\vp\times\vL+m^{2}\,V(r)\,\vx\right)\cdot\vn\,.
\eeq
In the previous expression, the mass ``$\,m\,$'' is preserved by the ``vertical'' reduction. Thus, in the Kepler case, we, therefore, deduce that on the dimensionally reduced flat 3-manifold, the symmetric tensor
\beq
C^{ij} =  2\,\delta^{ij}\,n_{k}\,x^{k}-n^{i}\,x^{j}-n^{j}\,x^{i}
\eeq
is a Killing-St\"ackel tensor generating the Runge-Lenz-type conserved quantity along the projection of the null geodesic (on the 5-manifold) onto the 3-manifold carrying the flat Euclidean metric.

\section{Conclusion}
In this paper, we studied the classical geodesic motion of a particle in Kaluza-Klein monopole space, and 
in its generalization, the Gibbons-Hawking space. We derived the equations of motion on the curved $3$-manifold, involving a curvature term, by viewing the conserved electric charge ``$\,q\,$'' as the momentum conjugated with the vertical coordinate on the extended manifold. In the original Kepler problem, the fixed mass ``$\,m\,$'' could be seen as associated with a translational symmetry generated by this fourth direction. 

We constructed, in particular, conserved quantities that are polynomial in the momenta by using the van Holten's recipe based on Killing tensors \cite{vHolten}. In section $\ref{KillingSection}$, we have discussed the conditions on Killing tensors which are related to the existence of constants of motion on the dimensionally reduced curved manifold. We have observed, in section $\ref{LiftedKillingTensor}$, that the Killing tensor generating the Runge-Lenz-type quantity preserved by the geodesic motion can be lifted to an extended manifold, namely, (\ref{LKT1}) and (\ref{LlightKillingTensor2}) \cite{DGH}. To illustrate our method, we have treated, in detail,  the generalized Taub-NUT metric, for which we derived the most general additional scalar potential so that the combined system admits a Runge-Lenz vector \cite{GW}.
It is worth noting that our formalism can be extended to spinning particles in Taub-NUT space \cite{VVHH}. Another example considered is the multicenter metric where we have found a conserved Runge-Lenz-type scalar (\ref{RLScalar}), in the special case of motions confined on a particular $2$-sphere. Moreover, Theorem \ref{RLenzCond} implies that,  for $N>2$, no Runge-Lenz vector does exist in the case of $N$-center metrics. 

In Ref.\cite{Plyushchay} it was observed that the Runge-Lenz-type vector plays a role also in supersymmetry. We also noted that apart from the generic importance of constructing constants of motion, namely in the confinement of particle to conic sections; the existence, in particular, of quadratic conserved quantity yields the separability of the Hamilton-Jacobi equation for the generalized Taub-NUT metric and for the two-center case which is relevant for diatomic molecules  \cite{GR,Nerses,Valent}.

\begin{acknowledgments}\noindent JPN is indebted to the {\it R\'egion Centre} for a doctoral scholarship and to the {\it Laboratoire 
de Math\'ematiques et de Physique Th\'eorique} of Tours University  for hospitality extended to him.
\end{acknowledgments}


\end{document}